**Tunability of the THz space-charge modulation in a vacuum microdiode**

P. Jonsson[1], Marjan Ilkov[1], A. Manolescu[1], A. Pedersen[2], A. Valfells[1,a]

[1]*School of Science and Engineering, Reykjavík University, 101 Reykjavík, Iceland*

[2]*Science Institute, University of Iceland, 101 Reykjavík Iceland*

[a]*Corresponding author*

Under certain conditions, space-charge limited emission in vacuum microdiodes manifests as clearly defined bunches of charge with a regular size and interval. The frequency correpsonding to this interval is in the Terahertz range. In this computational study it is demonstrated that, for a range of parameters, conducive to generating THz frequency oscillations, the frequency is dependant only on the cold cathode electric field and on the emitter area. For a planar micro-diode of given dimension, the modulation frequency can be easily tuned simply by varying the applied potential. Simulations of the microdiode are done for 84 different combinations of emitter area, applied voltage and gap spacing, using a molecular dynamics based code with exact Coulomb interaction between all electrons in the vacuum gap, which is of the order 100. It is found, for a fixed emitter area, that the frequency of the pulse train is solely dependent on the vacuum electric field in the diode, described by a simple power law. It is also found that, for a fixed value of the electric field, the frequency increases with diminishing size of the emitting spot on the cathode. Some observations are made on the spectral quality, and how it is affected by the gap spacing in the diode and the initial velocity of the electrons.

# I. Introduction

The study of space-charge limited emission has a long history, beginning with the classical Child-Langmuir law describing the maximum current density that can be drawn from a planar diode[1]. Other research of interest has included, e.g. the effects of geometry[2-4], finite emission energy[5], quantum effects[6], and temporal effects[7]. Space-charge limited emission is closely related to virtual cathode formation and has been used a source of radiation in devices such as the vircator[8], which is based upon virtual cathode formation and oscillation in a drift tube.

This study is a continuation of earlier work by the authors where a mechanism for bunching of electron beams at the cathode in a vacuum microdiode was discovered through numerical modelling[9]. This bunching occurs in situations where an unlimited supply of electrons is available for emission at the cathode, with negligible emission velocity, in the presence of a constant vacuum field for acceleration. Simulations show that initially a burst of electrons is emitted into the diode until the space-charge of the emitted electrons is sufficient to cause field reversal at the cathode surface, thereby inhibiting further emission. Typically this will happen well before the electron beam can fill the gap, i.e. the pulse length is shorter than the transit time. Once the electron bunch has moved sufficiently far from the cathode, or been absorbed at the anode, the electric field at the cathode reenters a state where emission is favorable, at which time a new bunch forms until field reversal at the cathode occurs again. This process can rapidly evolve into a periodic state characterized by bunches of electrons being emitted from the cathode at regular time intervals. These bunches will degrade over the propagation time due to internal forces, but if the gap spacing of the diode is appropriate, they can reach the anode with their structure and spacing largely intact.

Figure 1 shows the numbers of electrons per time-step emitted at the cathode and absorbed at the anode as a function of time in a 0.5µm diode with an applied voltage of 1V. Note the phase shift in the current due to the transit time. Another notable feature is the greater magnitude of the leading pulse in the pulse train.

The aforementioned work on space-charge modulation in microdiodes[9] describes the basic mechanism behind the origin of the bunching and obtains an approximate formula for the frequency, based on a simplified model of the system. In this work the same computational approach is used to carry out a systematic parametric survey to study the frequency spectra associated with the electron bunches as a function of gap spacing and applied potential in a planar diode for different values of the emitter area. From this survey a simple empirical power law is constrcuted, which relates the modulation frequency to the applied electric field for different radii of the emission area. These laws have the advantage of being based on the more exact physics of the simulation, as compared to the more qualitative relation for the frequency proposed in our earlier work[9]. This is of considerable importance for the purpose of designing a physical experiment to look for this type of modulation.

We will begin by presenting the physical model and simulation approach applied, followed by an exposition and a discussion of the main results.

## II. Model and simulation approach

The system under study is a parallel plate vacuum microdiode of infinite area. However, electrons are only emitted from a circular disk on the surface of the cathode with the radius $R_E$. In a Cartesian coordinate system the normal direction on the cathode and on the anode is the z axis, the two electrodes being located at z = 0 and z = D respectivly. Note that D is also the gap spacing.

It is assumed that there is an infinite supply of electrons present inside the cathode, beneath the cathode-vacuum interface, ready to be injected into the gap if the surface electric field is favorable (i. e. with a negative z component, leading to acceleration of the electrons away from the cathode, towards the anode). Thus emission can never be source-limited. Furthermore it is assumed that the electrons are emitted with negligible velocity and that they are point-like.

The dynamical behavior of the electrons is obtained by applying a molecular-dynamics method. The repulsive interactions between electrons present in the gap are modelled using an exact Coloumb potential for point charges. To drive the electrons from the cathode to the anode an external field is applied by setting the potential across the diode as $\Phi$. The time is discretized in small intervals $\Delta t$. An iteration in the simulation is carried out in three stages: A) Emission, B) Particle advancement and C) Absorption. We will now describe each component briefly.

### A. Emission

Particle emission is carried out using a Monte Carlo approach. In each time-step electrons are injected at random positions on the cathode surface. The injection is automatically suppressed when space-charge limited conditions are reached, namely when there is no spot on the cathode where the total electric field (which is the sum of the field created by the applied voltage and of the field created by the electrons present in the gap) is oriented in the negative z direction, so that it cannot transfer more electrons into the gap. The computational method is the following: A random point, $p_1 = (x_1,y_1,0)$, on the surface of the cathode within the emitter region $x^2 + y^2 < R_E^2$ is selected. To determine if this point is blocked by the present space-charge distribution a test is done. This test relies on the z-component of the accelerating field calculated at a point, $p_{1\alpha} = (x_1,y_1,-h/2)$, immediately below $p_1$, by adding the contribution from the applied vacuum field and the field from each of the N electrons already present in the diode gap. If the electric field at $p_{1\alpha}$ is favorable, then an electron is placed at the point, $p_{1\beta} = (x_1,y_1,h/2)$, which is directly above $p_{1\alpha}$, but seperated by a distance of h. Thus and the number of electrons in the gap becomes N + 1. In our simulations we give h the value of 2nm. The implications of picking this value will be discussed later. If the field at $p_{1\alpha}$ is not favorable then no placement of an electron occurs and a failure of injection is regitered. The next step is to pick another point, $p_{2\alpha} = (x_2,y_2,0)$, in the same random manner. Subsequently, the z-component of the electric field is calculated at $p_{2\alpha} = (x_2,y_2,-$

h/2). If the field is favorable for emission at that point an electron is placed at $p_{2\beta}$ = $(x_2,y_2,h/2)$, the number of electrons in the gap is increased by one, and we carry on picking points of emission at random. Similarly, if the field is unfavorable we place no electron, register a failure of injection and move on to pick another point. However, if there is a failure to inject in 100 consecutive attempts, we deem that space-charge limited conditions have been reached and cease trying to place more electrons. At this juncture the simulation moves on to the next step which is particle advancement.

## B. Particle advancement

At a given time, $t = t_i = i \times \Delta t$, there are $N_e$ particles in the gap. Here $\Delta t$ is the time step, which is equal to 1 fs in our simulations. The position of a particle at this time is given by $\mathbf{r}_{ni}$ = $(x_{ni}, y_{ni}, z_{ni})$, where n ranges from 1 to $N_e$. The force acting upon the particle at this time is $\mathbf{F}_{ni}$ = $(F_{xni}, F_{yni}, F_{zni})$, and is the sum of the force due to the applied vacuum field and the respective electrostatic force between electron n and all the other $N_e$-1 electrons as calculated from their positions at time $t_i$. We then determine the location for that particle at time $t_{i+1} = t_i + \Delta t$ using the Störmer-Verlet method

$$x_{n(i+1)} = 2x_{ni} - x_{n(i-1)} + \frac{F_{xni}}{m_e}(\Delta t)^2 \qquad (1)$$

where $m_e$ is the mass of the electron. The same method is used to calculate $y_{n(i+1)}$ and $z_{n(i+1)}$. For the electrons injected at time $t_i$ there is no defined position at time $t_{i-1}$ and in this case we use:

$$x_{n(i+1)} = x_{ni} + \frac{F_{xni}}{2m_e}(\Delta t)^2 \quad . \qquad (2)$$

The position of electrons are calculated in this manner as long as they are in the gap. Once they cross the x-y plane at either z = 0 or z = D, corresponding to the cathode and anode respectively, they are removed from the simulation. It is thus possible that a recently injected electron is pushed back into the cathode due to the evolving electrostatic field created by the space-charge distribution. This kind of event is however rare, the regular case being the propagation towards the anode.

## C. Absorption

For the purpose of this study, the interest lies in examining the number of electrons emitted from the cathode and absorbed by the anode as a function of time. At each time step the number of particles emitted at the cathode and absorbed at the anode is recorded. The frequency of absorption at the cathode is negligible compared to emission and absorption at the cathode. Typically there are up to 10 electrons either emitted or absorbed per time step.

Due to the discrete time step there is an inherent graininess in the absorption profile. In addition there is also the question of the immediacy of electron absorption, i.e. how well one may specify when an electron is absorbed. Consider an electron at the anode. Its kinetic energy is equal to the applied potential, and the corresponding deBroglie length is given by

$$\lambda_e = \frac{h}{\sqrt{2 m_e q_e \Phi}} \qquad (3)$$

where $h$ is Planck's constant, $m_e$ the mass of the electron as before, $q_e$ the elementary charge and $\Phi$ the applied potential. Near the anode, the velocity of the electron is close to its terminal velocity and one can easily estimate the time that it takes an electron, drifting at this velocity, to travel one deBroglie length. Let us call that time, $t_e$, then we have:

$$t_e = \frac{h}{2 q_e \Phi} \qquad (4)$$

For instance, if $\Phi = 1$ V we have $t_e \approx 2$ fs. To alleviate the discrete character of the simulation we incorporate a procedure after gathering the absorption data, whereby the charge of each electron is spread through time using a Gaussian filter. To wit, if $T = [T_1, \ldots, T_P]$ is a record of all the times an absorption event takes place in a given run of the simulation, then the Gaussian filter is applied to give a continuous function I(t) which describes the current at the anode at any time:

$$I(t) = \frac{q_e}{\sqrt{2\pi}\sigma} \sum_{k=1}^{P} \exp\left[-(t - T_k)^2 / 2\sigma^2\right] \qquad (5)$$

where σ is the width of the filter and has units of time. The same type of smoothing can be used to define a continuous current at the cathode. For the purpose of analysis the width of the filter should be picked so that $t_e < \sigma \ll T$, where T is a characteristic time describing the interval between bunches in the beam. A larger value of σ yields smoother current profiles and will also act as a high frequency filter in spectral analysis.

Figure 2 shows the effects of filtering: The raw data as presented as the number of absorption events at the anode per iteration, and the current at the anode obtained by passing the raw data through the Gaussian filter is shown as a dashed line. In this case σ = 150 time steps and T is approximately 1500 time steps.

### III. Results

Simulations were run for 64 combinations of applied potential and gap size, with a fixed circular emitting area of 250 nm radius. The values for the potential applied were $\Phi$ = [0.5, 1.0, 1.5, 2.0, 2.5, 3.0, 3.5, 4.0] V and the values for the gap size were $D$ = [0.5, 1.0, 1.5, 2.0, 2.5, 3.0, 3.5, 4.0] μm. Furthermore, 14 combinations of $\Phi$ and D were inspected for an emitter area with a radius of 100 nm, and for 6 combinations of $\Phi$ and D when the emitter radius was decreased to 50 nm. The choice of parameters was based on our previous work where a gap spacing of 1μm and applied potential of 1V with an emitter radius of 250 nmgave a frequency around 0.8 THz, indicating that this was the regime of interest. Since this type of modulation does not appear in large diodes we restricted our investigation to gaps smaller than 4 μm. Similarly gaps smaller than 500 nm were ignored, as quantum effects may become important at smaller length scales.

For each combination, time series recording the number of electrons emitted at the cathode and absorbed at the anode were generated. Fourier analysis was carried out for both the raw timeseries and also for the conduction current at the electrodes detemined by the above described Gaussian filtering. From the resulting spectra the dominant frequency as well as the full-width-at-half-maximum (FWHM) of the corresponding spectral peak are defined and have been determined.

Figure 3 shows the frequency spectrum for the anode current in the case of an applied voltage $\Phi = 1$ V and gap spacing of D = 0.5 μm for both unfiltered and filtered data. As can be seen the filtering process removes some of the high-frequency components, but does not alter the main bunching frequency.

We find that the dominant frequency, which describes the bunching, is a simple function of the applied electric field. This can be seen clearly in Figure 4. In fact, the relation between the frequency and the applied field can be described by a simple power law:

$$f = A \times E^\alpha \qquad (6)$$

where $f$ is the frequency measured in Hz, and $E$ is the applied vacuum field measured in V/m. The paremeters $A$ and $\alpha$ depend upon the size of the emitter, i.e. the radius of the circular are of emission on the cathode. Table I show the values of the parameters A and α for three inspected emitter radii.However, the width of the corresponding spectral peak is not so simply dependant on the applied field. The normalized FWHM -defined as the FWHM of the peak divided by the peak frequency- ranges from about 0.1 to 0.3, and is noticeably dependant on both the applied potential and gap size, rather than being dependant solely on the applied field.

**IV. Discussion**

The relation given in equation (6) shows a different scaling between frequency and field strength from the $E^{3/4}$ scaling predicted by the toy model that was originally used to explain the origin of space-charge modulation observed in the microdiode[9]. This difference can be explained qualitatively by noting that in the toy model a rigid disk shape was used to describe the bunches parallel to the emission plane, and more importantly only a single bunch was assumed to be present in the diode gap at any time. With more than one bunch in the gap, the trailing bunch(es) will be expected to experience a weaker accelerating field and therefore a longer time to travel the be sufficient distance from the cathode so that it may reopen for emission. This results in a longer interval between emission cascades and, thus in a lower frequency. The toy model also indicates that the frequency should rise with diminishing emitter radius. This effect can be seen clearly in Figure 4. The simple power law relation between the frequency and applied electric field, for fixed emitter areas, is a new and useful result, as it implies that a microdiode could be used as an easily tunable oscillator, and gives a quantitative means of predicting the frequency. Also, by selecting an appropriate radius for the emitter area it is possible to set the frequency range available for a given range of applied electric field strength.

The normalized FWHM is a measure of the quality of the pulse train as a possible source of radiation. We find that at the cathode the beam bunches are very distinct, but degrade as they propagate towards the anode due to internal repulsion between the electrons that make up each bunch. The degree of degradation is dependant on the magnitude of this internal force, and the time over which it can act. For a pulse of a given dimension, the magnitude of the force is increasing with the number of electrons in the pulse. The maximum number of electrons present in the gap of a planar diode under space-charge limited conditions can be shown to be directly proportional to the applied vacuum field by multiplying the classical Child-Langmuir current and transit time:

$$N_e \approx 100 \frac{\Phi[V]}{D[\mu m]} A[\mu m^2] \tag{7}$$

where $\Phi$ is the gap potential in V, D the gap spacing (measured in μm in this instance) and A the area of the emitter measured in μm$^2$. Thus the number of electrons per bunch is roughly proportional to the applied field. However, if the gap spacing increases while the applied field is kept constant, the propagation time will increase, leaving more time for the strong internal repulsive Coulomb forces to act. This can lead to the degradation of the pulse. For constant applied field the spread is more pronounced for higher potential and gap size. Ultimately, this explains why bunching of this sort can only be seen on the microscale. In macroscopic diodes the train of bunches will simply have washed out and become a continuous current. A competing effect is that of the number of bunches present within the gap. We have observed that for a fixed value of the applied field, the number of bunches within the gap increases with gap size, increasing from one pulse up to 4 or 5 pulses. The presence of this additional space charge sharpens the pulses emanating from the cathode, and conceivably inhibits spreading out of a bunch that is flanked on either side by other bunches that provide a focusing effect. Figure 5 shows how the normalized FWHM decreases with increasing potential, in the case where the applied field is constant, for 3 different values of the applied field. This is likely due to the effect of multiple bunches within the gap.

An obvious question arises in light of degradation of the pulses: Namely, what are the effects of the emission velocity? The results presented from parametric survey are obtained under the assumption that the emission velocity is zero. The authors did inspect two cases where effects caused by finite emission velocity were included. For these studies the longitudinal initial velocity of each individual emitted electron was assigned from a uniform distribution of emission energy ranging from 0 to $E_{max}$, where $E_{max}$ is a proportion of the terminal energy of electron striking the anode. In the first case $E_{max}$ was set equal to 5% of the terminal energy, resulting in so strong degradation of the pulses that no clear frequency was apparent and no distinct pulses remained. In the second case, $E_{max}$ was decreased to 1% of the terminal energy, and for this case a peak in the frequency spectrum for the anode current remained clearly intact. If one attributes the spread in emission velocity to thermal velocity of the electrons, then one can estimate the temperatures corresponding to the energy spreads used. For $\Phi = 1$ V we have $E_{max} = 0.01$ eV, roughly corresponding to a cathode temperature of 116 K. For $E_{max} = 0.05$ eV, the corresponding temperature is 580 K. Despite the fact that these estimates are based on a uniform rather than Maxwellian energy distribution, they do indicate that a microdiode would have to be operated at cooled conditions to generate bunches. Further study of the effects of the emission velocity will be presented in a seperate paper.

A few remarks on the method of emission are also in order. The reader should recall that once a point on the cathode surface has been selected from a uniform distribution, the longitudinal component of the electric field at this point but 1 nm behind the cathode surface is determined. If an electron can enter into the system at the inspected point will be placed 1 nm above the cathode. A possible objection to this method is that, for a typical time step of 1 fs, the distance of 1 nm is very large for the given field strengths, which are on the order of 1 MV/m. For these field strengths and the typical time step one might expect the electron to travel a distance of less than 1 pm, whereas the time needed to advance by 1 nm is of the order of 100 fs. This means that the intial spacing between injected electrons and the cathode is overestimated, and the retarding of the electric field, due to the space charge, might be underestimated. Our approach is however justified by several considerations: First, a real cathode will not be perfectly planar, and a length scale on the order of 1 nm to represent irregularity of the surface is not unreasonable. Second, if one takes into account a finite emission energy on the order of 1% of the terminal energy, or roughly 0.01 eV, we see that an electron with the corresponding velocity would travel a distance of 0.1 nm in 1 fs and have an associated deBroglie wavelength of the order 100 nm, thus the 1 nm length scale falls in between the two aforementioned in a natural manner. Finally, and perhaps most importantly, one should consider the longitudinal component of electric field due to an electron that has been placed above the cathode. Figure 6 shows an example for illustration: An electron (shown as a solid dot) is already in place at a distance of h/2 directly above the cathode surface and a point (shown as an open circle) has been picked for testing the field according to our algorithm for injection. This point is at a lateral distance d from the present electron and h/2 below the cathode. The contribution of the pre-existant electron to the longitudinal electric field at the point of testing can be expressed as:

$$E_z = \frac{-eh}{4\pi\varepsilon_0(h^2+d^2)^{3/2}} \qquad (8)$$

which is a function of both h, and d. Thus the longitudinal field due to the pre-existing electron will be strongly affected by d, as can be seen in Figure 8, which shows the longitudinal field as a function of d for three different values of h. Refering back to equation (7) it is evident that for A of the order of 1 μm$^2$, Φ of the order of 1 V, and D of the order of 1 μm, the number of electrons in the gap is about 100 and the number of electrons scattered across the emitter area is no greater than 100. Thus the lateral spacing between electrons, d, will typically be much greater than 10 nm. Bearing this in mind, we may see from Figure 7 that, in this case higher values of h will yield higher estimates of the retarding field, and subsequently lower estimates of the number of electrons per pulse, which is an observed effect in our simulations. By varying h, we see some variation in the number of electrons per pulse, but it does not affect the bunching mechanism.

## V. Conclusion

A parametric survey examining space charge modulations has been conducted for planar vacuum microdiodes with gap lengths in the range of 0.5 – 4 μm and applied voltages ranging from 0.5 – 4 V. The modulation of the anode current is typically of the order of 1 THz and can be described with a simple power law given in equation (6). The exact shape of the curve relating frequency and applied field is dependant on the radius of the circular emitting area on the cathode. Thus, in principle, it should be possible to design a microdiode oscillator that can be tuned by varying the applied DC electric field, with the frequency range being set by the emitter size. The normalized width of the peak in the spectrum associated with the modulation is a complex function of appled potential and gap size. For a given value of the applied field the frequency spectrum of the anode current seems to initially become sharper with increasing gap size, but ultimately degrades as the gap size becomes to large. First indications are that a spread in electron emission energy corresponding to 1% of the energy at the anode will not disrupt the modulation.

**Acknowledgements**

This work was supported by the Icelandic Research Fund grant number 120009021. The authors would also like to thank Dr. John Verboncoeur for useful discussions.

| Emitter radius [nm] | A | α |
|---|---|---|
| 50 | 779×10$^6$ | 0.539 |
| 100 | 326×10$^6$ | 0.580 |
| 250 | 257×10$^6$ | 0.575 |

**Table I.** Magnitude of the parameters in Equation (6), $f = AE^\alpha$, for different values of the radius of the emitter area.

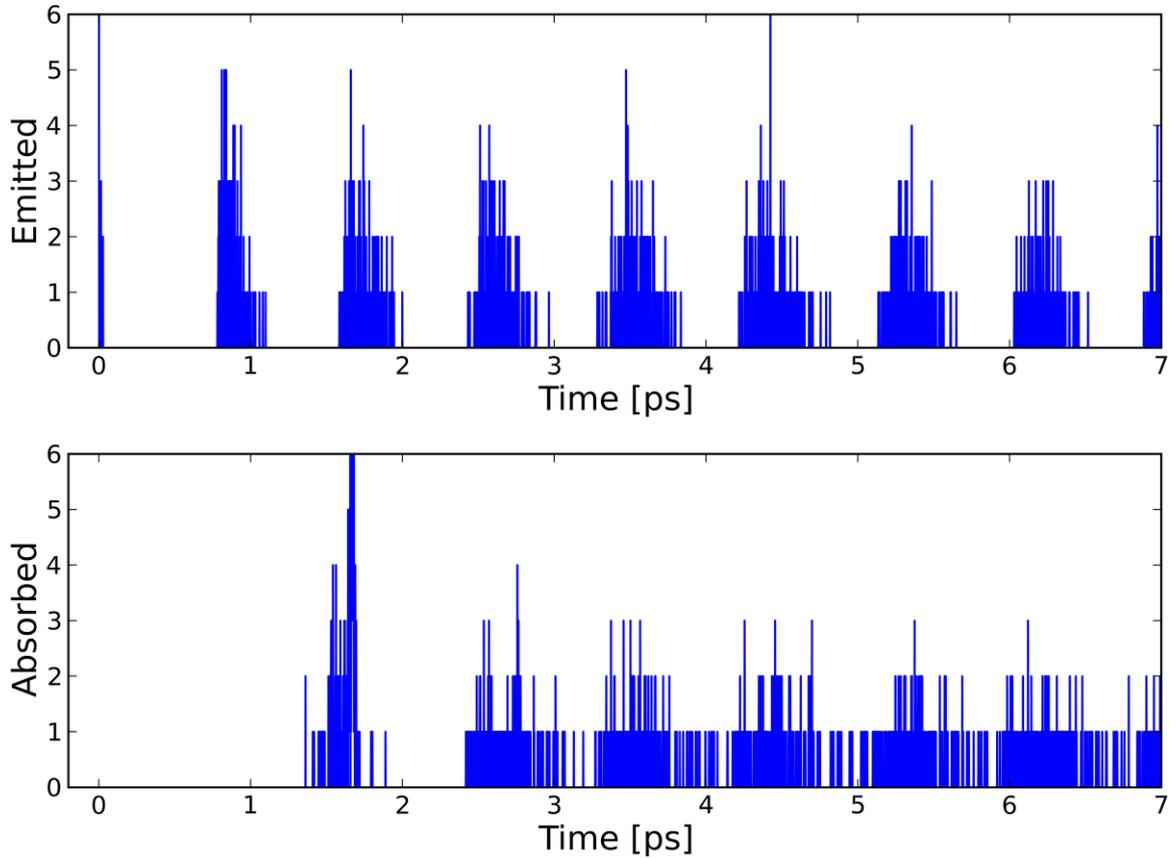

**Figure 1.** Number of electrons per time step emitted at the cathode (upper) and absorbed at the anode (lower) in a planar diode with gap spacing of 0.5µm and applied potential of 1V. The radius of the emitting area is 250 nm.

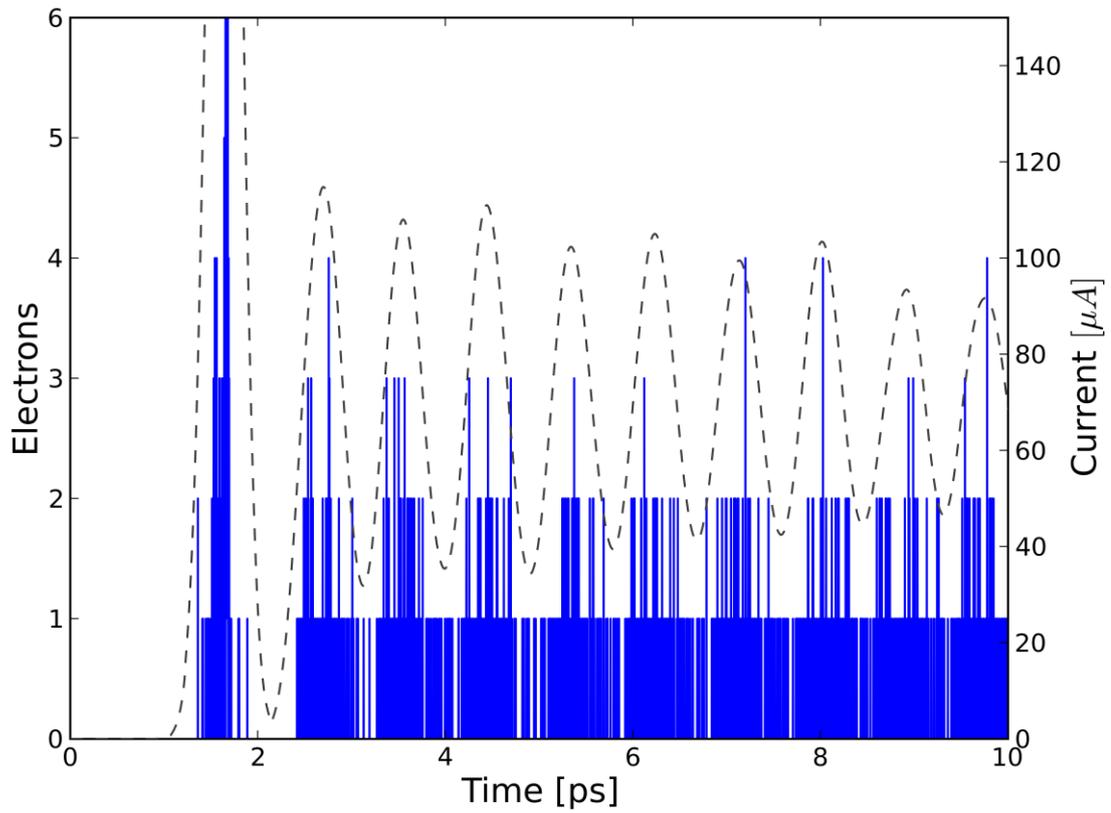

**Figure 2.** Comparison of raw data showing number of electrons absorbed at anode per iteration and current at anode obtained using Gaussian filter (dashed line). The same operating parameters are used as in Figure 1, while the Gaussian smoothing filter uses σ = 150 fs.

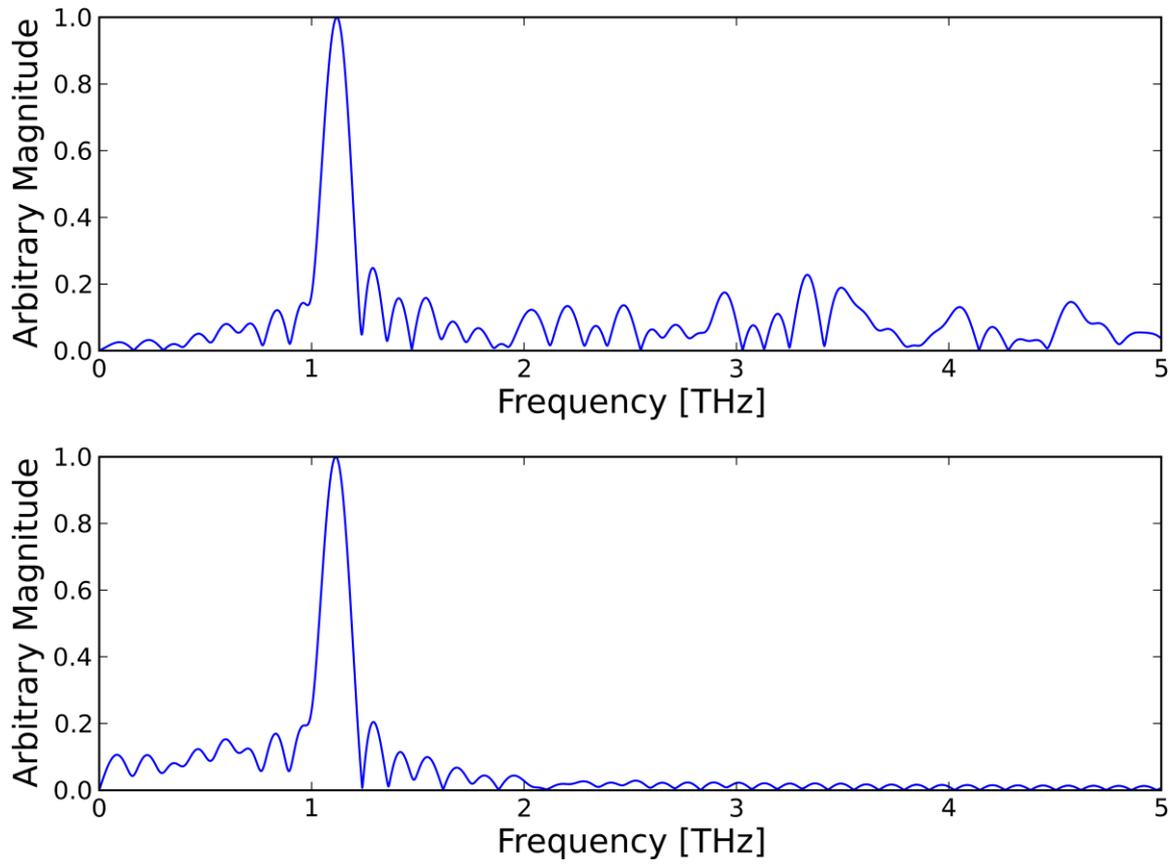

**Figure 3.** Spectrum of raw data for electron absorption at anode (upper) and anode current obtained using Gaussian filter (lower). The same parameters are used as in Figure 2.

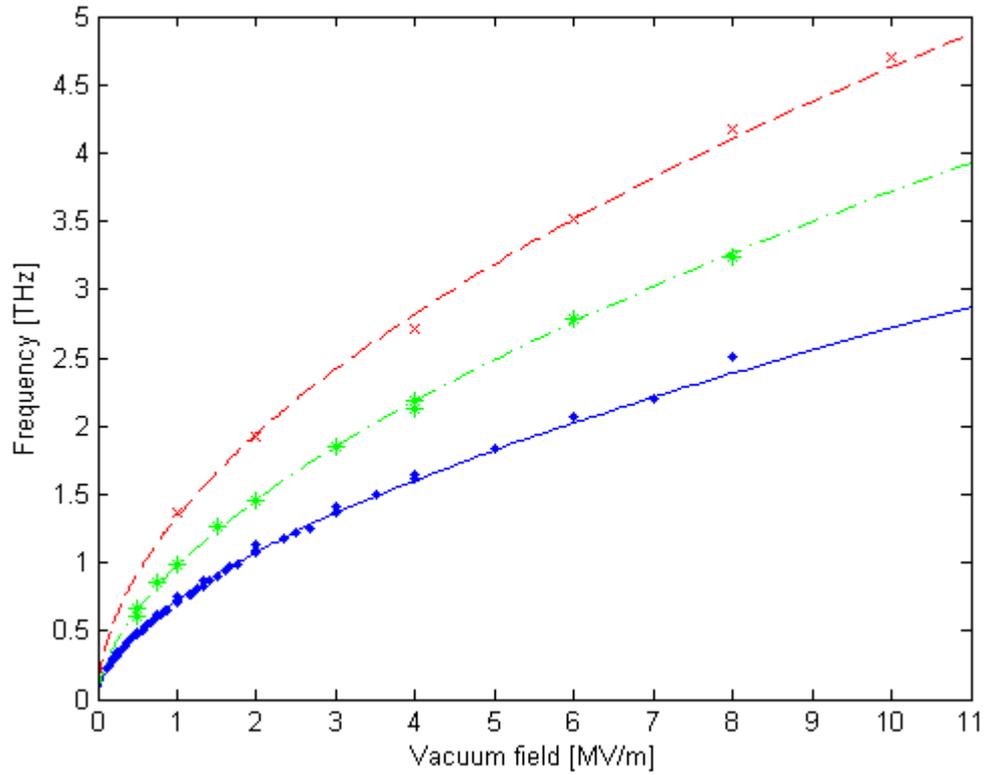

**Figure 4.** Modulation frequency as a function of applied vacuum field shown for 84 different combinations of gap size, applied potential and emitter size. The solid line represents the frequency as described by Equation (6) fitted to the data (dots) for $R_E$ = 250 nm. The dash-dotted line represents the fitted to the data (stars) for $R_E$ = 100 nm. The dashed line represents the fitted to the data (crosses) for $R_E$ = 50 nm.

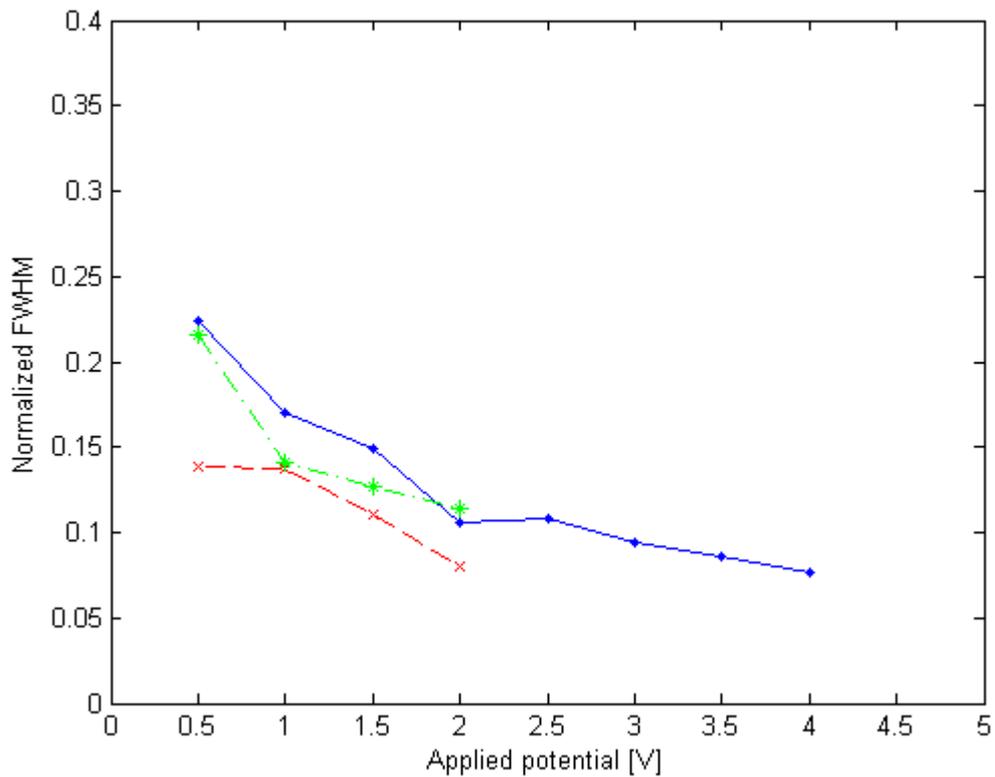

**Figure 5.** Normalized FWHM of the spectral peak associated with the modulation frequency as a function of the applied potential, shown for a fixed value of the applied vacuum field of 1MV/m (dots and solid line); 0.5 MV/m (crosses and dashed line); and 2 MV/m (stars and dash-dotted line). The emitter radius is 250 nm.

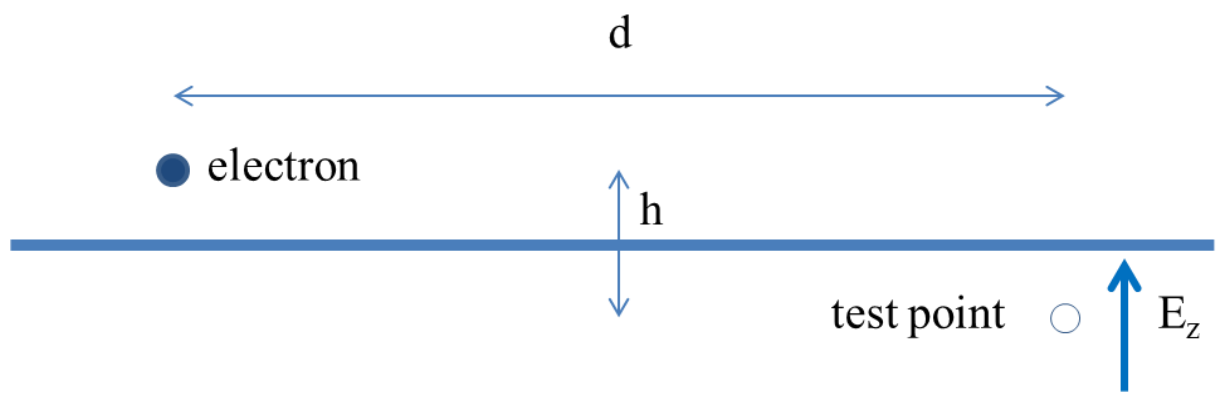

**Figure 6.** Schematic of example to examine the longitudinal electric field, in the test plane under a possible injection point, due to the presence of a nearby electron in the placement plane.

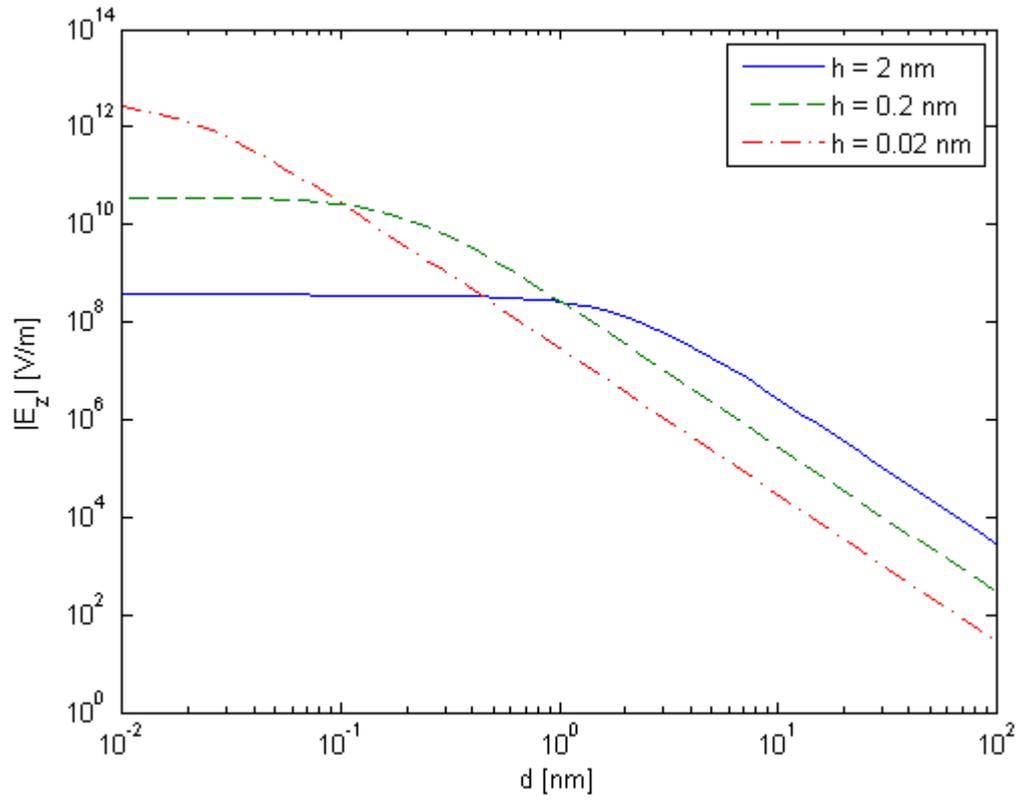

**Figure 7.** Magnitude of the longitudinal electric field at the test point due to an electron in the placement plane as a function of d and h.